\documentclass{article}

\begin{document}
\thispagestyle{empty}

\begin{flushright}
JLAB-THY-01-11\\
20 April 2001
\end{flushright}

\begin{center} 
{\bf\Large
 QCD Calculations of Pion Electromagnetic and Transition
Form Factors}\footnote{To appear in the proceedings of 3rd 
Workshop on Chiral Dynamics - Chiral Dynamics 2000: 
Theory and Experiment, Newport News, Virginia,
17-22 Jul 2000.}  \\[1cm] 
\end{center} 

\begin{center} 
{\sc A.V. RADYUSHKIN}\footnote{Also at Laboratory of Theoretical
Physics, JINR, Dubna, Russia}  
\\[2mm] 
{\em Physics Department, Old Dominion University,} \\ 
{\em Norfolk, VA 23529, USA} 
\\[2mm] {\em and } \\[2mm] 
{\em Theory Group, Jefferson Lab,} \\ 
{\em Newport News, VA 23606, USA}
\end{center}
%

The transition  $\gamma \gamma^* \to \pi^0$
in which a  real and a  virtual photon produce 
a  pion  is the cleanest
exclusive process  for testing QCD predictions.
In the lowest order of
perturbative QCD, the asymptotic behavior of the relevant form factor 
$F_{\gamma \gamma^* \pi^0} \left(Q^2\right)$
is given by \cite{Lepage:1980fj} 
\begin{equation}
F_{\gamma^* \gamma^*  \pi^0 }^{LO}(Q^2) = \frac{4\pi}{3}
\int_0^1 dx{{\varphi_{\pi}(x)}\over{xQ^2}} \,   +
O(\alpha_s/\pi) + O(1/Q^4)\  , 
\label{eq:g*g*pipqcd}
\end{equation}
where $\varphi_{\pi}(x)$ is the pion distribution amplitude 
describing the projection  of the pion onto the quark-antiquark
pair carrying the momenta  $xp$ and $(1-x)p$.  
The  nonperturbative information is accumulated here by
the integral
\begin{equation}
I_0 = \int_0^1 {{\varphi_{\pi}(x)}\over{x}} dx \
\label{eq:I }  .
\end{equation}
Its value depends on the shape of $\varphi_{\pi}(x)$.
In particular,  using  the
asymptotic form \cite{pl80,Lepage:1980fj}
$
\varphi_{\pi}^{as}(x) = 6 f_{\pi} x(1-x)
$
one obtains $I_0=3 f_{\pi}$ which gives 
the $
F_{\gamma \gamma^*  \pi^0 }^{as}(Q^2) = {4 \pi f_{\pi}}/{Q^2} $ 
prediction for the large-$Q^2$
behavior \cite{Lepage:1980fj}.
Brodsky and Lepage \cite{Lepage:1980fj}  
 proposed the   interpolation formula
 $F_{\gamma \gamma^*  \pi^0 }^{BL}(Q^2)
 = 
 1 / [\pi f_{\pi} 
 (1+Q^2/s_0)] 
 $
which reproduces, for $s_0 = 4 \pi^2 f_{\pi}^2$,  
both the $Q^2 =0 $ value 
and the high-$Q^2$ behavior.   
The same result follows from the model \cite{acta} 
based on local quark-hadron duality,
in which $F_{\gamma \gamma^*  \pi^0 }(Q^2)$ 
is obtained by calculating  the 
amplitude of the transition of the 
pion into a $\bar q q$ pair with the invariant
mass $s$, with subsequent integration 
over the pion duality interval $0<s<s_0$.

Adding the one-loop pQCD corrections \cite{ggpirc,murad}  
and assuming the asymptotic distribution amplitude (DA) one obtains 
\begin{equation}
F_{\gamma \gamma^*  \pi^0 }^{NLO}(Q^2)\left |_{\varphi=\varphi^as} 
 = \frac{4 \pi f_{\pi}}{Q^2} \right. 
\left \{ 1 - \frac{5}{3} \,  {{\alpha_s}\over{ \pi}}\right \}  \  .
\label{19}  \end{equation}
Another  frequently used model $
\varphi_{\pi}^{CZ}(x) = 6 f_{\pi} x(1-x)(1-2 x)^2
$ was proposed by Chernyak and Zhitnitsky \cite{cz}.
It gives  an essentially larger value  $I_0=5f_{\pi}$.
Comparison with recent CLEO data \cite{CLEO} favors 
the distribution amplitudes close to the asymptotic one.

The asymptotic behavior of the pion electromagnetic 
form factor can be also calculated in pQCD \cite{cz,pl80,Lepage:1980fj}
\begin{equation}
F_{  \pi }^{LO}(Q^2) = \frac{8\pi \alpha_s (Q^2)}{9}
\int_0^1 dx \int_0^1 dy \, {{\varphi_{\pi}(x)\varphi_{\pi}(y)}\over{xyQ^2}} =
\frac{ 8\pi \alpha_s (Q^2)}{9 Q^2}  I_0^2 \  . 
\label{eq:pipqcd}
\end{equation}
It involves the same integral $I_0$ of the pion DA.
However, taking $I_0 \approx 3 f_{\pi}$ (as suggested 
by the $\gamma \gamma^*  \pi^0$ data) and 
 $\alpha_s \sim 0.3$  gives the result which 
 is too small. It is instructive to rewrite  the pQCD term 
 (for  the asymptotic DA)    as $2(\alpha_s/\pi)(s_0/Q^2)$.
 Since $s_0=4 \pi^2 f_{\pi}^2 \approx 0.67$ GeV$^2 \sim m_{\rho}^2$,
 the pQCD  term has an extra  
 factor $2(\alpha_s/\pi) \sim 0.2$ compared 
 to the  $m_{\rho}^2/Q^2$ behavior suggested 
 by the VMD model $ F_{  \pi }^{VMD}(Q^2) =1/(1+Q^2/m_{\rho}^2)$. 
 Note, that  the $O(\alpha_s/\pi)$ factors are the standard 
 penalty  for each extra loop in Feynman diagrams. 
 Hence, the natural way out is to add 
the  contribution which has the zeroth 
order in $\alpha_s$. It corresponds to overlap
of the soft parts of the pion wave functions.
This nonperturbative contribution
can be estimated using the local 
quark-hadron duality model. 
Calculating the  amplitude for the transition
$\bar q q \gamma^* \to \bar q' q'$,
with the initial  $\bar q q $ pair having 
mass $s_1$ while  the final $\bar q' q' $ 
pair having mass $s_2$, and integrating 
over the pion duality region $0<s_1,s_2<s_0$ 
one obtains \cite{nerad} 
$$
F_{\pi}^{LD,soft}(Q^2) = 1 - \frac{1+6s_0/Q^2}{(1+4s_0/Q^2)^{3/2}} \  . 
$$
Asymptotically, this contribution decreases as $1/Q^4$.  
Within the local duality approach,
the $\alpha_s/Q^2$ term is obtained from the
$O(\alpha_s)$ contribution to the $\bar q q \gamma^* \to \bar q' q'$ 
amplitude.  Fortunately, the $Q^2 =0$ limit of this contribution
is fixed by the Ward identity resulting
in $F_{\pi}^{LD,\alpha_s}(Q^2=0) = \alpha_s / \pi$. 
The simple interpolation 
formula   (analogous to the Brodsky-Lepage 
expression) gives
$$
F_{\pi}^{LD,\alpha_s}(Q^2) = 
\left (\frac{\alpha_s }{ \pi} \right ) \frac1{1+Q^2/2s_0} \  . 
$$
The sum $F_{\pi}^{LD,soft}(Q^2) + F_{\pi}^{LD,\alpha_s}(Q^2)$ 
is in a full agreement (for $\alpha_s / \pi =0.1$) 
with the recent Jefferson Lab data \cite{Volmer:2001ek}.  
Similar results for the pion form factor have been obtained 
within the light-cone QCD sum rule approach \cite{Braun:2000uj}.

The pQCD radiative corrections to the asymptotic term
are known \cite{radcorr}.
In case of the asymptotic DA, the result in the $\overline{MS}$
scheme is \cite{radcorr,Brodsky:1998dh,Bakulev:2000uh}
$$
F_{\pi}^{pQCD,NLO}(Q^2) = \frac{8 \pi f_{\pi}^2 \alpha_s (Q^2)  }{ Q^2} 
\left \{ 1+ \frac{\alpha_s }{ \pi} 
\left [ \frac76 \beta_0 - 3.91 \right ] \right \} \  , 
$$
where  $\beta_0 =11 - 2N_f/3$ 
is the lowest coefficient of the QCD $\beta$ function.
The $O(\beta_0)$ term can be  absorbed into the 
redefinition of the argument of the QCD running  
coupling $\alpha_s (Q^2) \to \alpha_s (Q^2e^{-14/3})$, which 
indicates  that the average virtuality of the exchanged
``hard'' gluon is much smaller than $Q^2$. 
Numerically, for all accessible $Q^2$,   
the scale $Q^2 e^{-14/3}$ is well below 
the typical hadronic scales like $m_{\rho}^2$,
so one should treat $\alpha_s (Q^2e^{-14/3})$ 
as an effective constant $\sim 0.4$ corresponding to $\alpha_s$
taken in the  ``infrared'' limit, below which 
$\alpha_s$ does not run \cite{Brodsky:1998dh}. 
The remaining negative correction has the same nature
as the  $O(\alpha_s)$ term in the expression 
for the $\gamma \gamma^* \to \pi^0$ form factor. 
They both  are  due to the Sudakov effects \cite{murad}
which sqeeze the effective transverse size of 
$\bar q q$ pairs \cite{Li:1992nu}.

Summarizing, both the perturbative and nonperturbative
aspects of the $Q^2$ dependence of the 
$\gamma \gamma^* \to \pi^0$ form factor and of the pion
electromagnetic form factor 
are  rather well understood in quantum chromodynamics.
However, new experimental data at higher $Q^2$
would be extremely helpful for  detailed  
tests of the transition to the regime 
where the pQCD hard contribution plays the dominant role.

This work was supported by the U.S. 
Department of Energy under contract
DE-AC05-84ER40150 under which the Southeastern
Universities Research Association (SURA)
operates the Thomas Jefferson National Accelerator
Facility (Jefferson Lab).

\end{document}